\documentclass[aps,amsmath,twocolumn,a4paper,showpacs]{revtex4}

\usepackage{graphicx}
\usepackage[dvips]{color}
\usepackage{mathrsfs}
\usepackage{amssymb}

\newcommand{\tensadd}{\sigma_{\rm add}}
\newcommand{\tenstot}{\sigma}
\newcommand{\tensel}{J}

\begin{document}

%%%%%%%%%%%%%%%%%%%%%%%%%%%%%%%%%%%%%%%%%%%%%%%%%%%%%%%%%%%%%%%%%%%%%
\title{Elastic moderation of intrinsically applied tension in lipid membranes}

\author{Michael A. Lomholt}
\affiliation{MEMPHYS - Center for Biomembrane Physics, Department of Physics and Chemistry, University of Southern Denmark, Campusvej 55, 5230 Odense M, Denmark}
\author{Bastien Loubet}
\affiliation{MEMPHYS - Center for Biomembrane Physics, Department of Physics and Chemistry, University of Southern Denmark, Campusvej 55, 5230 Odense M, Denmark}
\author{John H. Ipsen}
\affiliation{MEMPHYS - Center for Biomembrane Physics, Department of Physics and Chemistry, University of Southern Denmark, Campusvej 55, 5230 Odense M, Denmark}

\date{\today}

%%%%%%%%%%%%%%%%%%%%%%%%%%%%%%%%%%%%%%%%%%%%%%%%%%%%%%%%%%%%%%%%%%%%%

% A B S T R A C T

\begin{abstract}
Tension in lipid membranes is often controlled externally, by pulling
on the boundary of the membrane or changing osmotic pressure across a
curved membrane. But modifications of the tension can also be induced in
an internal fashion, for instance as a byproduct of changing a membranes
electric potential or, as
observed experimentally, by activity of membrane proteins. Here we develop
a theory which demonstrate how such internal contributions to the tension
are moderated through elastic stretching of the membrane when the membrane
is initially in a low tension floppy state.
\end{abstract}


\maketitle

%%%%%%%%%%%%%%%%%%%%%%%%%%%%%%%%%%%%%%%%%%%%%%%%%%%%%%%%%%%%%%%%%%%%%

\section{Introduction}
Lipid membranes are ubiquitous in biological systems. Their primary function is to separate cells and organelles from their surrounding environments, but they also participate actively in many biological processes. For instance nerve conduction, exocytosis and endocytosis, and production of ATP \cite{alberts02}.

An important mechanical quantity for membranes is tension. For instance
endocytosis can be regulated by changes in tension \cite{dai97}. Often
the tension is controlled by external means. For example in micropipette
experiments, where a vesicle is aspirated to a pipette, the tension is set
through Laplace's law by the pressure difference between the inside and the
outside of the pipette \cite{kwok81}. Using Langmuir-Blodgett troughs the
tension in lipid monolayers on a water surface can be lowered by compression
\cite{israelachvili85}.
Or red blood cells can burst by increasing osmotic pressure
when the induced tension exceeds the tensile strength. However,
there are also intrinsic ways for the tension of a membrane to be modified
when the membrane is attached to a fixed frame (or encloses a fixed
volume). Different scenarios for this case are: (i) Changes in an electric
membrane potential, say when an action potential travels along a nerve, will
induce a change in the tension \cite{lacoste07,ambjornsson07}. (ii) For the
force-dipole model introduced in \cite{manneville01} to explain micropipette experiments
on membranes with active ion pumps, it was argued in \cite{lomholt06_mech} that the model
implies an additional active contribution to the tension of the membrane. (iii)
If a membrane becomes charged (say due to a change of pH) then this will also
lead to a contribution to the tension \cite{chou97,kumaran01}. (iv) If the membrane exchanges material with a reservoir, say lysolipids with the surrounding bulk fluid \cite{needham95}, then the tension of the membrane is also likely to be affected. However, this situation will be more complicated than the previous three cases. For instance, the overall equilibration of the system will depend on balance equations for the chemical potentials of the lysolipids in and out of solution. We will not consider this case further in the following.

For each of the situations (i)-(iii) theoretical predictions have been
made for the additional contribution that the effect will add to the tension. The
question we will address in this paper is: how will the total tension of the
membrane respond to such contributions. The question has gained
actuality due to recent experiments on the fluctuation spectrum of active
membranes \cite{faris09}. In these experiments a decrease in tension from
typically $4\times 10^{-7}$ N/m to $0.5\times 10^{-7}$ N/m was observed when
ion pumps in the membrane were activated. The magnitude of this change is much
smaller than the a priori contribution $\sim$$10^{-3}\,$N/m predicted by the
force-dipole model of the membrane protein activity \cite{manneville01,lomholt06_mech}. From
the vast discrepancy between these numbers one might be lead to conclude
that a more complicated force
distribution than a dipole must be used to explain the data. However, in
\cite{lomholt06_mech} arguments were sketched indicating that the a priori
contribution would not necessarily be equal to the total change in tension.
The purpose of the present paper is to present systematic
arguments following the formalism of \cite{fournier01,henriksen04}
to show how elastic contributions will moderate the effect of intrinsically
applied tension. We will limit ourselves to situations in which only the tension is modified, and not other parameters such as bending rigidity, compressibility etc.

The paper is organized as follows. We will first present our model in Section
\ref{sec:II}. Then we will calculate fluctuations and the behavior of the tension in
Section \ref{sec:III}. In Section \ref{sec:IV} we will discuss the consequences
for the different
examples of intrinsic tensions mentioned above.
Finally we give conclusions
and outlook in Section \ref{sec:V}.

\section{The Model}\label{sec:II}
As our starting point we will take a Hamiltonian with three contributions
\begin{equation}
\mathcal{H}=\mathcal{H}_c+\mathcal{H}_s+\mathcal{H}_{\rm add}.\label{eq1}
\end{equation}
The first contribution is the Helfrich bending energy \cite{helfrich73}
\begin{equation}
\mathcal{H}_c=\int dA\,\frac{\kappa}{2}(2 H)^2
\end{equation}
where the integral is over the area of the membrane and $H$ is the mean curvature at a specific point on the membrane. In Monge gauge, where the membrane is parametrized by the height $z=h(x,y)$ of the membrane above the $x$$y$-plane, we have to second order in $h$: $dA=dx\,dy\{1+[(\partial_x h)^2+(\partial_y h)^2]/2\}$ and $(2H)^2=(\partial_x^2 h+\partial_y^2 h)^2$.
The second contribution represents elastic stretching contributions. Expanding around a prefered area $A_0$ one can write
\begin{equation}
\mathcal{H}_s=\frac{K_a}{2 A_0}(A-A_0)^2
\end{equation}
where $A=\int dA\,1$ is the actual area of the membrane and $K_a$ is the area expansion modulus. These first two contributions have been found to describe well experiments on lipid vesicles, including micropipette experiments spanning a wide range of tensions \cite{fournier01,henriksen04}. We have not included a spontaneous curvature or related elastic contributions (as in the ADE model \cite{miao94}), since they will not matter for the discussion below (in the Monge gauge $H$ is a total derivative up to second order in $h$ and thus maximally contribute a boundary term). Similarly, the Gaussian curvature is not included since by the Gauss-Bonnet theorem it also only contributes a boundary term when the topology of the membrane is fixed. The final contribution in Eq. (\ref{eq1}) represents the additional intrinsically applied tension
\begin{equation}
\mathcal{H}_{\rm add}=\tensadd A.
\end{equation}
The question we will investigate in the following is: ``how does the intrinsically applied tension $\tensadd$ affect the fluctuation spectrum?'' 

An obstacle to answering this question is the nonlocality of $\mathcal{H}_s$. To get around it we will follow the approach of \cite{fournier01,henriksen04} and use the Hubbard-Stratonovich transformation to write ($\beta^{-1}=k_B T$ is Boltzmann's constant times temperature)
\begin{multline}
\exp\left[-\beta\frac{K_a}{2 A_0}\left(A-A_0\right)^2\right]=\sqrt{\frac{-\beta A_0}{2\pi K_a}}\\
\times \int_{i\infty}^{-i\infty}d \tensel\, \exp\left[\beta\frac{A_0}{2 K_a}\tensel^2-\beta \tensel\left(A-A_0\right)\right]
\end{multline}
meaning that we introduce a new auxiliary field, $\tensel$, and replace the Hamiltonian by
\begin{equation}
\mathcal{H}'=\mathcal{H}_c+\tensadd A-\frac{A_0}{2 K_a}\tensel^2+\tensel\left(A-A_0\right).
\end{equation}
We then take the thermodynamic limit and evaluate the integral over $\tensel$
using the method of steepest descent, i.e.,  we assume that fluctuations in $\tensel$ are small and can be neglected (for a justification see \cite{henriksen04}). Thus we let $\tensel$ take its stationary value obeying $\partial F/\partial \tensel=0$, where $F=-\beta^{-1}\ln {\rm Tr} \exp(-\beta \mathcal{H}')$ is the Helmholtz free energy. This gives
\begin{equation}
\tensel=\frac{K_a}{A_0}\left(\langle A\rangle -A_0\right).\label{eq:J}
\end{equation}
The notation introduced here is that for a quantity $X$ its thermal average is: $\langle X\rangle\equiv {\rm Tr} X \exp(-\beta \mathcal{H}')/Z\equiv \int [d h]\,X \exp(-\beta \mathcal{H}')/Z$, where $Z\equiv \int [d h]\, \exp(-\beta \mathcal{H}')$ is the partition function and $[d h]$ a measure for integrating over all possible shapes.

\section{Fluctuations and tension in Monge gauge}\label{sec:III}
To connect with the tension obtained in for instance a video microscopy experiment we need to find the magnitude of the fluctuations of the membrane shape. To calculate these we choose the Monge gauge introduced in the last section, and assume that the membrane is attached on a quadratic frame of side length $L$. Expanding in a Fourier series with coefficients $h_q=\int d x d y\,e^{-i(x q_x+y q_y)}h(x,y)$, where $q_x=2\pi n_x/L$ and $q_y=2\pi n_y/L$ with $n_x$ and $n_y$ integers, we have to second order in $h$ and its derivatives
\begin{eqnarray}
A&=&L^2+\frac{1}{2 L^2}\sum_{n_x,n_y} q^2|h_q|^2\label{eq8}\\
\mathcal{H}_c&=&\frac{\kappa}{2 L^2}\sum_{n_x,n_y} q^4 |h_q|^2
\end{eqnarray}
where $q=\sqrt{q_x^2+q_y^2}$. Using the equipartition theorem one finds within this Gaussian approximation the fluctuations in the shape to be
\begin{equation}
\langle |h_q|^2\rangle = \frac{k_B T L^2}{\kappa q^4 +(\tensadd+\tensel)q^2}\label{eq10}.
\end{equation}
Defining the tension $\tenstot$ as the tension that can be observed in video microscopy experiments of fluctuating vesicles, i.e., the coefficient in front of the $q^2$ term in the denominator of Eq. (\ref{eq10}) (as done for instance in the study of active membranes in \cite{faris09}) we have $\tenstot=\tensadd+J$. Inserting in Eq. (\ref{eq8}) we can obtain the excess area $\alpha$ of the membrane relative to the frame
\begin{equation}
\alpha\equiv\frac{\langle A\rangle -L^2}{L^2}=\frac{1}{L^2}\sum_{n_x,n_y}\frac{k_B T q^2/2}{\kappa q^4 +\tenstot q^2}
\end{equation}
To evaluate the sum over $n_x$ and $n_y$ we will assume that the tension lies within the interval $\kappa q_{\rm min}^2 \ll \tenstot \ll \kappa q_{\rm max}^2$, where $q_{\rm min}=2\pi/L$ while the upper cut-off $q_{\rm max}\sim 2\pi/(2 d)$ with $2 d$ being the membrane thickness. The lower bound corresponds to the experimental requirement that a tension dominated regime is observable in the fluctuation spectrum, while the upper bound will not be very far from the tensile strength of the membrane. With this assumption we can approximate $\sum_{n_x,n_y}\sim L^2\int_{q_{\rm min}}^{q_{\rm max}} q d q/(2\pi)$ and get \cite{REMalpha}
\begin{equation}
\alpha\sim\frac{k_B T}{8\pi \kappa}\ln \frac{\kappa q_{\rm max}^2}{\tenstot}.\label{eq12}
\end{equation}
Since fluctuations at high $q$ values are not affected by changes in tension we will simply assume that the cut-off $q_{\rm max}$ is also constant. Thus combining Eqs. (\ref{eq12}) and (\ref{eq:J}) we get
\begin{equation}
d\tenstot-d\tensadd=d \tensel=\frac{K_a L^2}{A_0}d\alpha=- \sigma_c \frac{d\tenstot}{\tenstot},
\end{equation}
where $\sigma_c=K_a\frac{L^2}{A_0}\frac{k_B T}{8\pi \kappa}$. Rearranging this we get a relationship between changes in the intrinsically applied tension $\tensadd$ and the observable tension $\tenstot$
\begin{equation}
\left(1+\frac{\sigma_c}{\tenstot}\right)d\tenstot=d\tensadd
\end{equation}
This is the main result of this paper. The formula can straightforwardly be integrated to give the additional contribution as a function of the observed tension and an integration constant which we take to be the tension $\sigma_{\rm ref}$ in the absence of the additional contribution
\begin{equation}
\tensadd=\tenstot-\sigma_{\rm ref}+\sigma_c\ln\frac{\tenstot}{\sigma_{\rm ref}}.
\end{equation}

If we restrict ourselves to the low tension limit $\tenstot \ll \sigma_c$ (but still having $\tenstot\gg \kappa q_{\rm min}^2$) then we can obtain the relation
\begin{equation}
\tenstot=\sigma_{\rm ref}\exp(\tensadd/\sigma_c).\label{eq16}
\end{equation}
In this limit the additional contribution will not affect the observed tension with its full effect. Instead it will be offset almost completely by a change of the membrane area $A$ leading to a countering elastic contribution $dJ$ to the observed tension. Note that we have used the assumption that $\tenstot\gg \kappa q_{\rm min}^2$ to derive Eq. (\ref{eq16}). Thus we should not expect Eq. (\ref{eq16}) to hold for arbitrarily large negative $\tensadd$. In particular, one should expect $\tenstot$ to become negative at sufficiently negative $\tensadd$. And the membrane will then become unstable if the tension reaches $\sigma<-\kappa q_{\rm min}^2$.

In the opposite limit $\tenstot \gg \sigma_c$ one simply has
\begin{equation}
d\tenstot=d\tensadd.
\end{equation}
In this limit the excess area is too small for the membrane to contract and offset the additional contribution to the tension. Thus the additional contribution will have its full effect on the observed tension.

\section{Discussion}\label{sec:IV}
From the results derived in the previous section we see that additional contributions to the tension will have to be comparable in size to $\sigma_c$ to have an observable effect on the fluctuation spectrum. If we take typical values $K_a\approx 0.2\,{\rm N}/{\rm m}$, $A_0\approx L^2$, $\kappa\approx 20 k_B T$ \cite{henriksen04} one gets $\sigma_c\approx 4\times 10^{-4}\,{\rm N}/{\rm m}$. Let us try to compare this value with the expected values for the first three intrinsic contributions mentioned in the introduction.

(i) If an electric potential $V_m$ is applied across a lipid membrane then it was found in \cite{lacoste07,ambjornsson07} that a negative contribution to the tension proportional to $V_m^2$ will be induced. For potentials of the order 50\,mV it was found that the contribution would be $\tensadd\sim -10^{-5}\,{\rm N/m}$ \cite{REM1}.
Thus with the above estimate for $\sigma_c$ this electric contribution to the tension should not be easily observable.

(ii) For the case of active membranes, then we see that the prediction of a
tension contribution of the order $\sim$$10^{-3}\,$N/m from the force-dipole
model for a membrane in a floppy state by Eq. (\ref{eq16}) results in a
change of tension corresponding to a couple of factors of $e$. Although this
estimate agrees well with the observations of \cite{faris09}, then the estimate
is highly uncertain. A major effect, which is being neglected when applying
Eq. (\ref{eq16}), is the factor two change in ``effective temperature''
(or more precisely a coefficient corresponding to the factor
$k_B T/(8\pi \kappa)$ in front of the logarithm in Eq. (\ref{eq12}))
measured for
this type of active membrane \cite{manneville99}.
The effect of the increased ``effective temperature'' on the tension should be significant
since the implied increase in fluctuations at short wavelengths will compete for
excess area with the fluctuations in the long wavelength tension dominated
regime \cite{REM2}. The theory developed here do not take into account such changes in
``effective temperature''. Thus the only conclusion that can be drawn for the
active membrane case is that one cannot immediately conclude from the large difference
between the additional contribution of $\sim$$10^{-3}\,$ N/m estimated in
\cite{lomholt06_mech} and the change in observed value of order $\sim$$10^{-7}\,$ N/m
reported in \cite{faris09} that the force-dipole model does not hold. Further investigations are necessary before we have an understanding
of how the activity modifies a basic physical property like tension for
lipid membranes.

(iii) Another case is if a membrane becomes charged, for instance due to a change in pH when some of the lipids are acidic. Electrostatic interactions will then generate a tension as calculated in \cite{chou97} for example. The additional contributions to the tension can easily reach a magnitude around $\sigma_c$ even if just one percent of the lipids becomes monovalently charged under physiological salt conditions. However, such a change in charge content will also alter the bending rigidity of the membrane significantly, breaking the assumption of unaltered bending rigidity made for the derivations in this paper.

\section{Conclusions and outlook}\label{sec:V}
The present paper studied changes in the tension observed from fluctuations when so called intrinsic contributions were applied. It was found that when the membrane was in a low tension state with excess area then an intrinsically applied tension would be almost fully cancelled by an elastic contribution arising from contraction of the membrane area.

A major limitation of the present work is that it does not consider situations with changes in bending rigidity, area expansion modulus or prefered area, not to mention situations with non-equilibrium activity leading to additional sources of random noise in the system. An extension to these situations would have to take into account how fluctuations change also at very short wavelengths, around the scale determined by the upper cut-off for the wavenumbers ($q_{\rm max}$).

We also did not discuss effects on the mechanical tension of the membrane (the tension obtained by differentiating the free energy of the membrane with respect to the area of the frame on which it is attached). There is still some controversy about the relation between mechanical tension and the tension observed from fluctuaions (see \cite{cai94,farago03,henriksen04,barbetta10}). We avoided this issue here by only considering a fixed frame area $L^2$.

\begin{acknowledgments}
We would like to thank Jonas Henriksen, Jean-Fran{\c c}ois Joanny, David Lacoste, Per Lyngs Hansen and Ole Mouritsen for stimulating discussions.
\end{acknowledgments}

\bibliography{refs}

\begin{thebibliography}{24}
\bibitem[{REM({\natexlab{a}})}]{REMalpha}
\bibinfo{note}{Note that the definition of $\alpha$ as excess area in this
  paper differs from a definition of $\alpha$ as change in observable area used
  in micropipette experiments. In particular, Eq. (\ref{eq12}) is still valid
  at high tension even though it does not contain directly a term with $K_a$ as
  in Eq. (3) of \cite{evans90}.}
\bibitem[{REM({\natexlab{b}})}]{REM1}
\bibinfo{note}{Note that there is a misprint in Fig. 3 of \cite{ambjornsson07}:
  the value of $\sigma_{\rm el}^0$ should be around minus a few
  $10^{-5}\,$N/m.}
\bibitem[{REM({\natexlab{c}})}]{REM2}
\bibinfo{note}{This effect of competition for the excess area was not taken
  into account in the estimate of the activity's effect on the tension given in
  \cite{lomholt06_mech}.}
\end{thebibliography}

\end{document}